\documentclass{desyproc}
\begin{document}


\title{Vector Mesons and DVCS at HERA}



%
%
%
%
%
%
%
%
%

\author{{\slshape Marcella Capua}  for the H1 and ZEUS Collaborations\\[1ex]
Universit\`a della Calabria and INFN, Arcavacata di Rende (CS), 87036, Italy}


\contribID{ZZ}
\confID{UU}
\desyproc{DESY-PROC-2012-YY}
\acronym{MPI@LHC 2011}
\doi
\maketitle


\begin{abstract}

We present a review of the latest exclusive diffraction results achieved at HERA, discussing in particular the $W$ and $t$ dependence of the measured cross-sections in both DIS and photo-production regimes. 

\end{abstract}



\section{Introduction}

Exclusive diffractive processes, $ep\rightarrow eXp$, where the  system $X$ is a vector meson (VM) or a real photon,
have been extensively studied by the H1 and ZEUS Collaborations at HERA~\cite{vm,dvcs,h1,io}.
These processes help to understand the transition from soft to hard Quantum Chromo-Dynamics (QCD) 
and to study hard diffraction at HERA at a large $\gamma p$ centre-of-mass energy $W$ with a variable hard scale provided by the photon virtuality $Q^2$. 

Hard diffraction can be described in terms of perturbative QCD, at the leading order, by the exchange of two partons with different longitudinal and transverse momenta in a colourless configuration. 
In addition, at the HERA energies, the measurement of the Deeply Virtual Compton Scattering (DVCS) cross-sections provide constraints on the generalised parton distributions (GPDs)\cite{gpd}. The dependence of the GPDs on the four-momentum transfer squared at the proton vertex, $t$, gives information on the transverse distribution of partons in the proton which is not accessible through the measurements of the $F_2$ structure function. 

In this contribution we will concentrate on the latest HERA results on exclusive diffraction discussing in particular the $W$ and $t$ dependence in both the Deeply Inelastic Scattering (DIS) and photo-production regimes ($Q^2<2$~GeV$^2$).

\section{$W$ dependence}

\begin{figure}[h]
\includegraphics[scale=0.5]{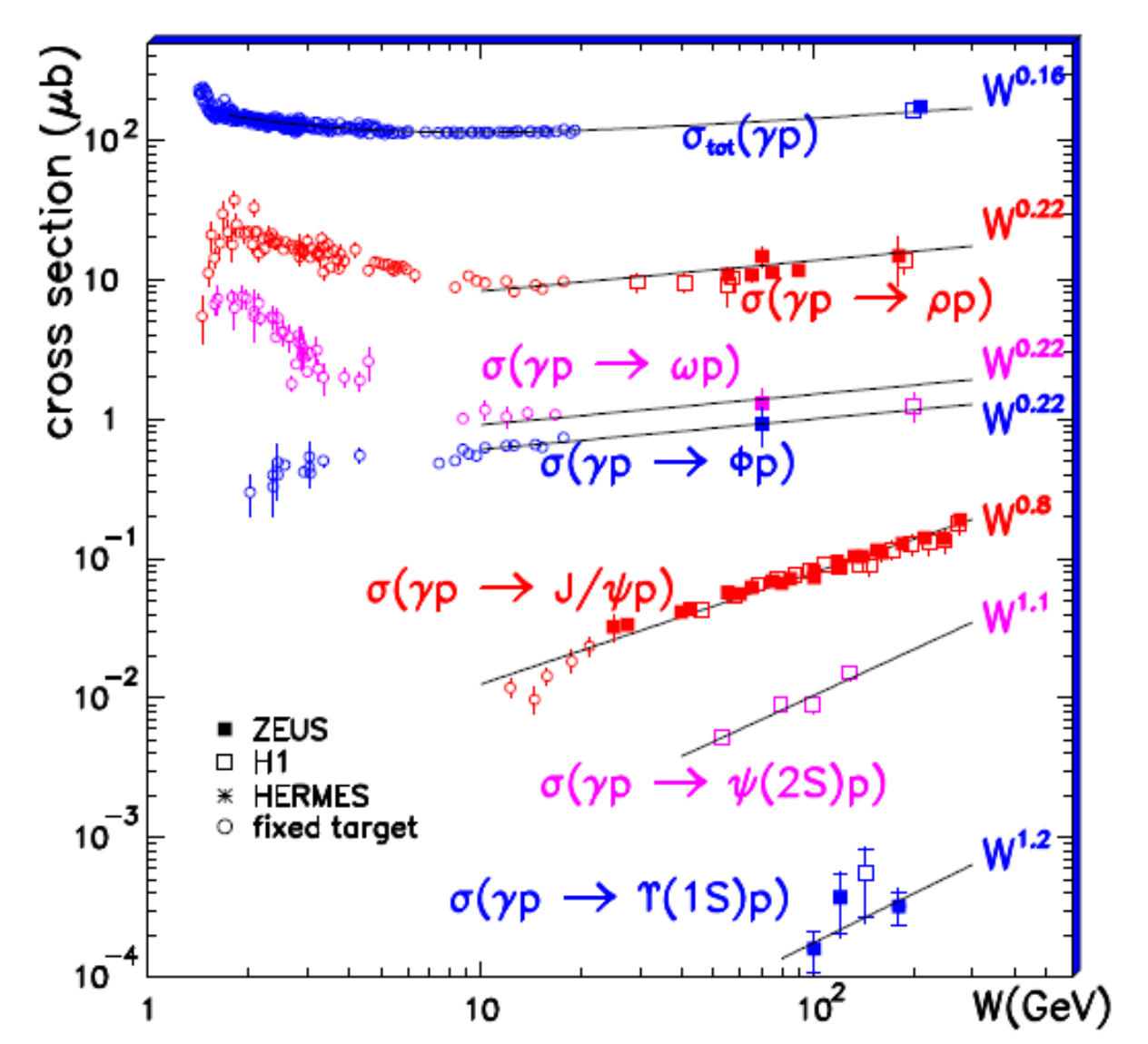}
\includegraphics[scale=.55]{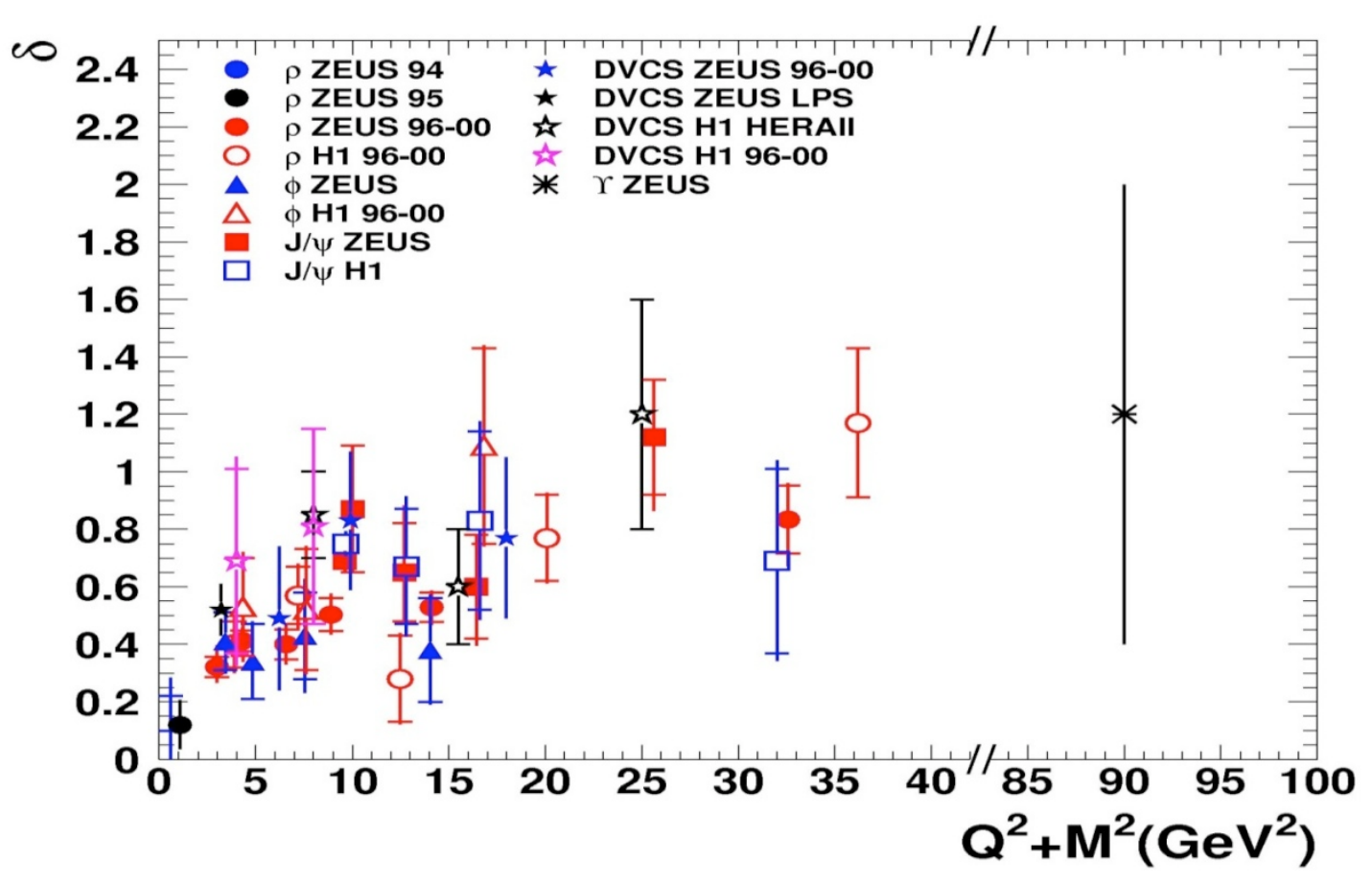}
\caption{(left) Compilation of the total and exclusive VM photo-production cross sections, as a function of $W$; the curves are the result of a fit of the form $W^{\delta}$. (right) Compilation of the parameter $\delta$,  as a function of the scale $(Q^2+M^2)$, obtained from the measured cross-sections for the exclusive VM in DIS, including  also DVCS and photo-production measurements.}
\label{fig1}
\end{figure}

\begin{figure}[h]
\centering
\includegraphics[scale=.47]{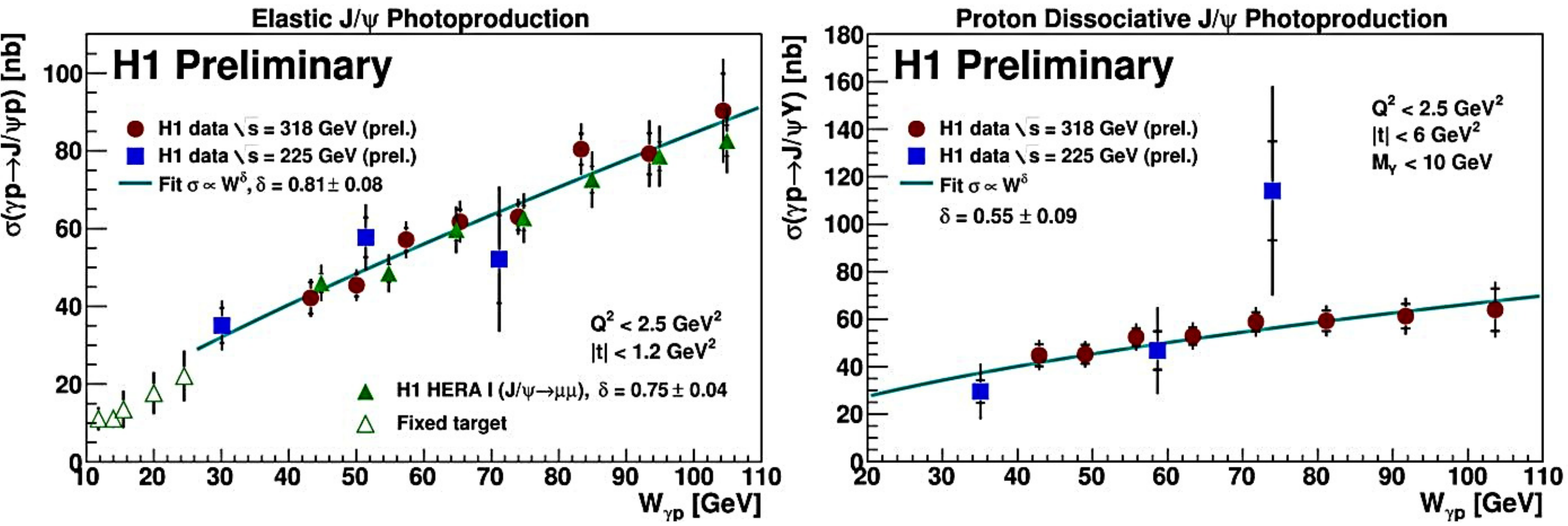}
\caption{$J/\psi$ photo-production cross sections as a function of $W$ for elastic (left) and proton dissociative measurements (right). The curves are fits of the form $W^{\delta}$.}
\label{fig2}
\end{figure}

 It is expected that the VM production cross section increases with increasing $W$ energy with a power law ($\sigma \sim W^{\delta}$), where the power factor $\delta$ grows with $Q^2$. Figure~\ref{fig1} (left) shows the total $\gamma p$ cross-section together with a compilation of light and heavy VM cross-section measurements in photo-production ($Q^2\sim 0$) as a function of $W$. The lines with slope $\delta$ guide the eyes. 
The W dependence of the total $\gamma p$ cross-section is typical of soft interactions and a similar behavior is visible also for light VMs. A different and stronger energy dependence can instead be observed for heavy VM (e.g. $J/\psi$) with the power $\delta$ increasing as the mass of the vector meson increases. This is interpreted as the evidence of gluon participation in the interaction, since the higher the scale at which the gluons are probed, the faster the rise with $W$.
The power $\delta$ is shown as a function of the scale $(Q^2+M^2)$ in fig.~\ref{fig1} (right), for a similar compilation of exclusive processes in DIS, including  also DVCS and photo-production measurements.

Still not included in fig.~\ref{fig1} (left) is the new preliminary H1 result on $J/\psi\rightarrow e^+e^-$ in photo-production~\cite{newjpsi}, which is shown in fig.~\ref{fig2}. This measurement is based on the last part of the HERA data taking, when three different proton beam energies were provided  (920 GeV, 575  GeV and 460 GeV). 
The different proton beam energies allow to extend the diffractive $J/\psi$ measurements towards lower $W$  energies. 

Figure~\ref{fig2} (left) shows the elastic $ep\rightarrow e J/\psi~p$ cross-sections as a function of $W$ compared with previous fixed targed and H1 measurements and the resulting value of the fit of the data to $W^{\delta}$ 
is in very good agreement with previous H1 and ZEUS (not shown in the figure) measurements. In the same figure (right) is presented the proton dissociative cross-section  $ep\rightarrow e J/\psi~ Y$ in which the proton dissociates into a low-mass state $Y$. The curve is the result of the fit to a power $\delta$. Can be noticed the different value of the slope $\delta$.

\section{$t$ dependence}

The $t$ dependence of the exclusive diffractive differential cross-section, $d\sigma /dt \propto e^{-bt}$, has been also investigated at HERA. We expect that because of the increasing hardness of the interaction, the $t$ distribution becomes universal, independent of the scale and the final state observed. The asymptotic value of $b$, the parameter of the exponential slope of the $t$ distribution,  reflects the size of the proton.

\begin{figure}
\centering
 \includegraphics[scale=0.45]{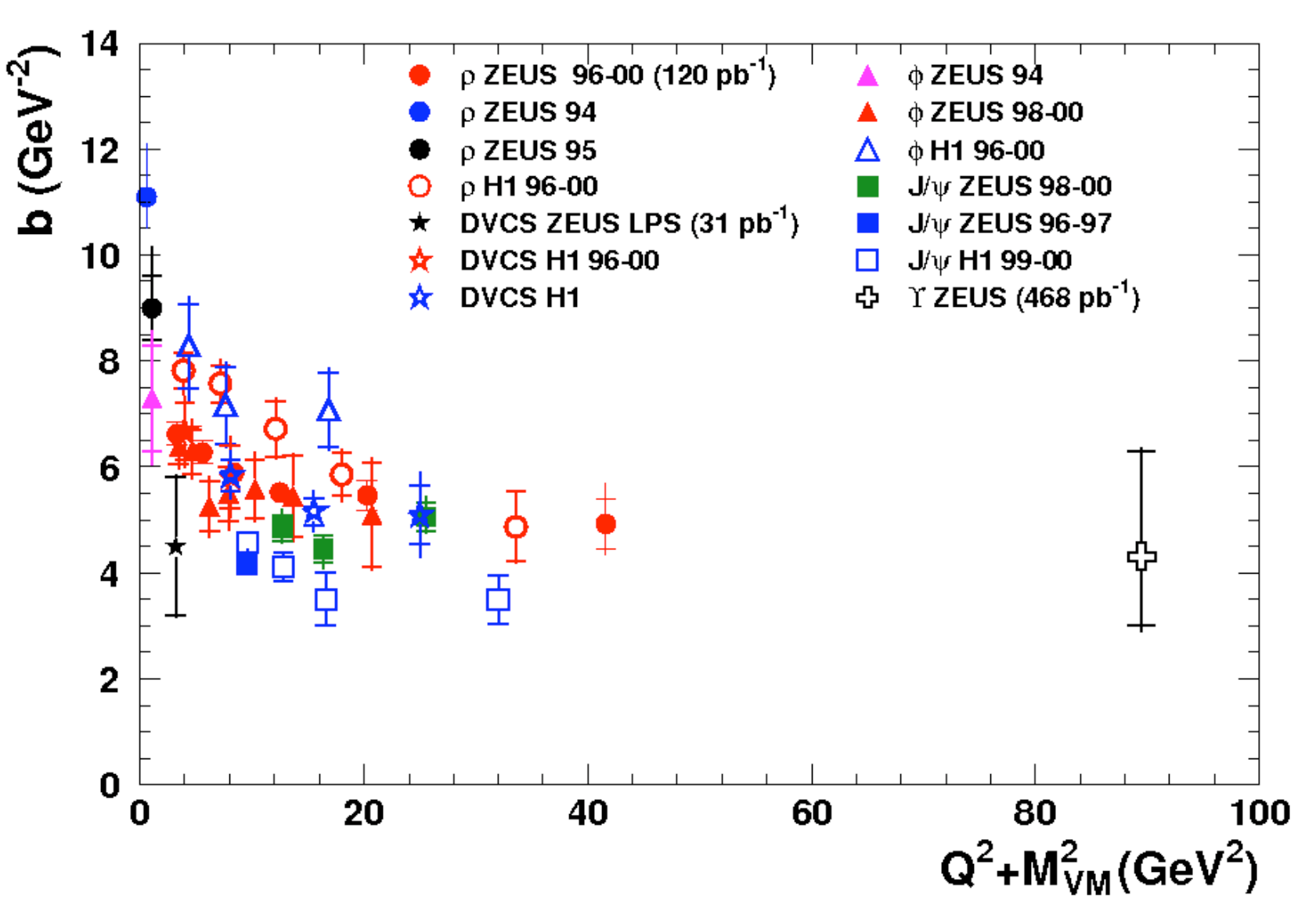}
\caption{Comparison of the HERA measurements of the slope parameter $b$  
as a function of the scale $Q^2 + M_{VM}^2$ for exclusive VM production and for DVCS.}
\label{fig3}
\end{figure}

Figure~\ref{fig3} shows a compilation of the $b$ slope values obtained from an exponential fit to the differential cross sections  $d\sigma /dt$ as a function of the scale $(Q^2+M^2)$ for various VMs, including the new ZEUS $\Upsilon(1S)$~\cite{tupsilon} measurement and recent DVCS  (for which $M=0$)~\cite{dvcs,h1,io} measurements.
A transition from the soft to the hard regime is visible, with $b$ decreasing with the increase of the scale to an asymptotic value close to 5 GeV$^{-2}$ which is an indication that the gluons are well contained within the proton. 
In fig.~\ref{fig3} the ZEUS DVCS measurement~\cite{io}, at $Q^2=3.2$~GeV$^2$,  has been obtained, for the first time, from the direct measurement of the $t$ variable with a dedicated spectrometer~\cite{lps}. In the H1 DVCS measurements~\cite{dvcs,h1}, the $t$ variable is computed as the vector sum of the transverse momenta of the final state photon and the scattered lepton; the kinematic region covered is $6.5<Q^2<80$~GeV$^2$, $30<W<140$~GeV and $t<1$~GeV$^2$. H1 has also published the differential cross section as a function of $t$ for different values of $Q^2$ and $W$ and observes a soft $Q^2$ dependence of the parameter $b$ while no  dependence has been observed as a function of $W$.

\begin{figure}
\centering
 \includegraphics[scale=.45]{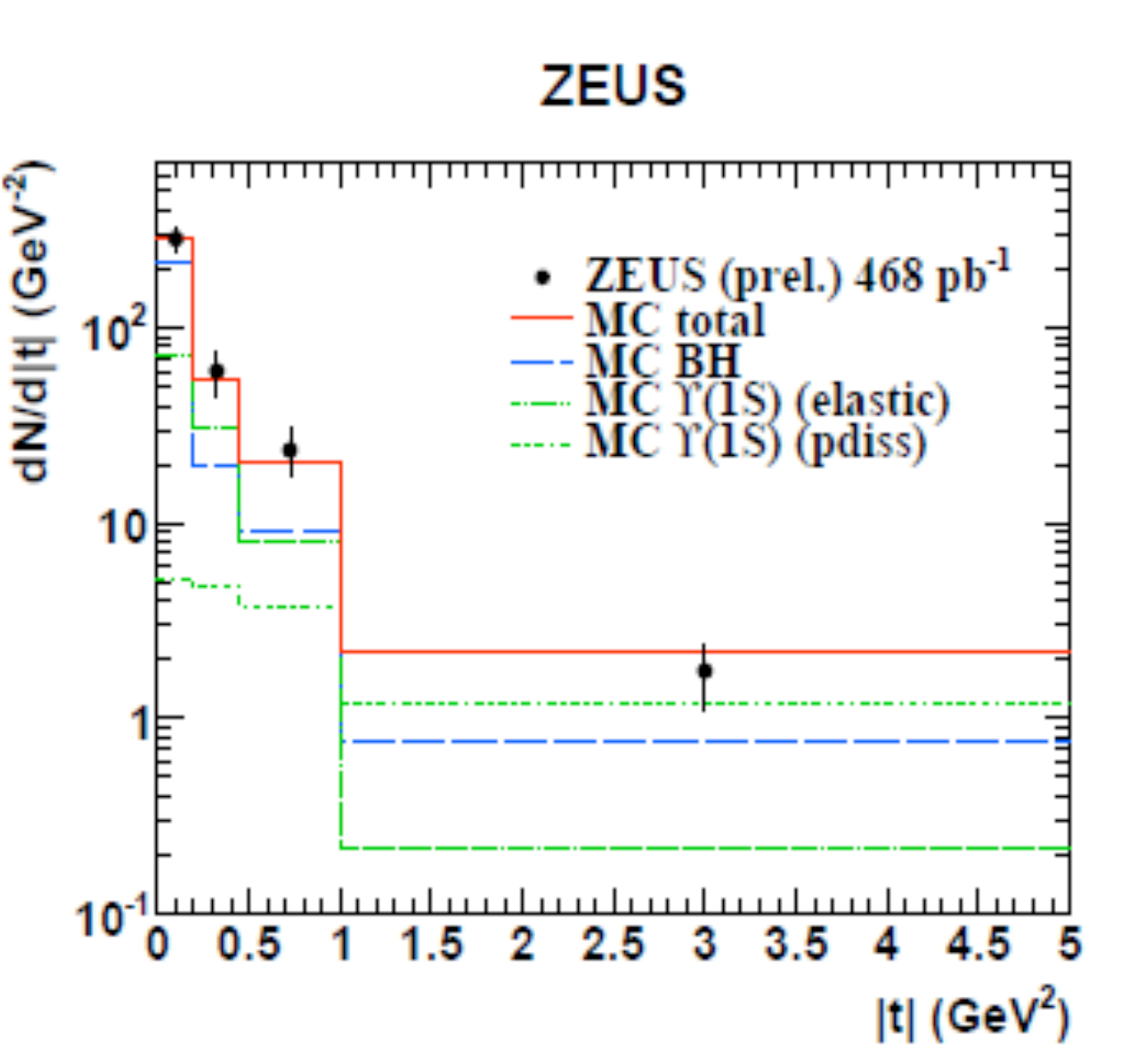}\\
   \includegraphics[scale=.5]{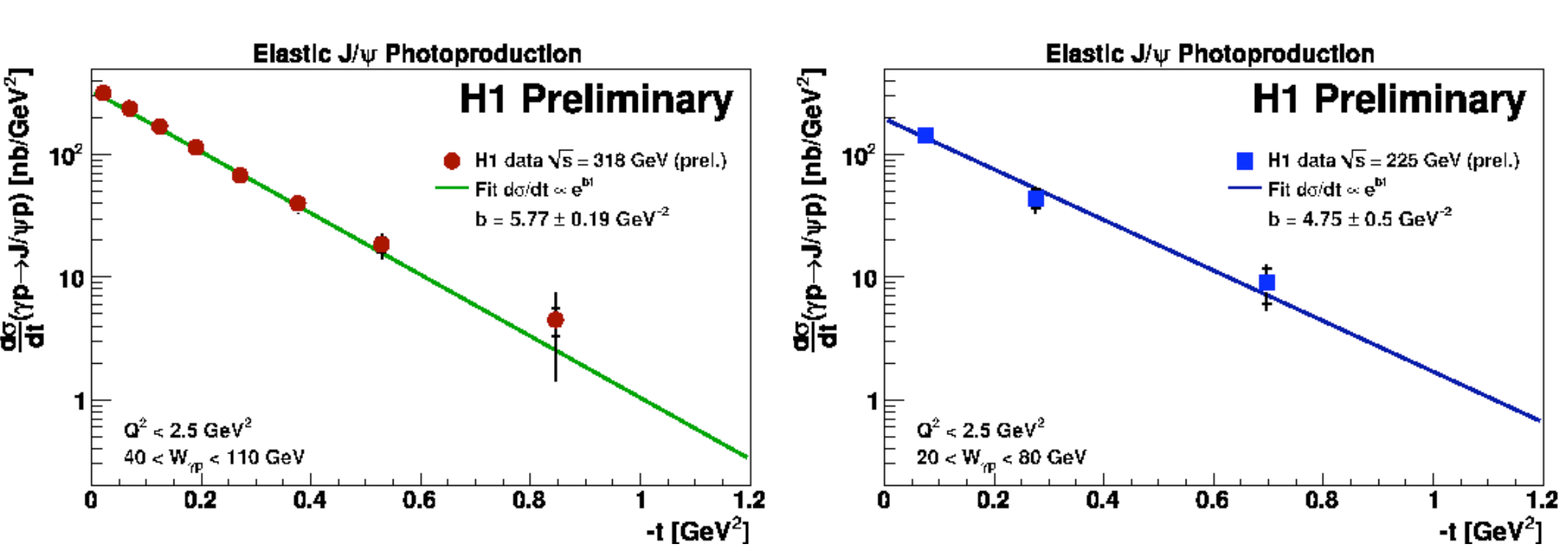}
\caption{(top) Measured $|t|$ distribution (full dots) with  error bars denoting statistical uncertainties. 
Fitted distributions for simulated events are shown for the Bethe-Heitler (dashed line), exclusive $\Upsilon(1S)$ (dotted line) and proton dissociative $\Upsilon(1S)$ (dashed-dotted line) processes.
The solid line shows the sum of all contributions. (bottom) Differential cross-section for exclusive $J/\psi$ photo-production as a function of $t$ for two proton beam energies. The lines are the results of an exponential fit.}
\label{fig4}
\end{figure}

In addition, Figure~\ref{fig3} shows the first determination of the $b$ parameter for $\Upsilon(1S)$ production.
The exclusive photo-production reaction $\gamma p \rightarrow \Upsilon(1S) p$ was studied by the 
ZEUS detector using the entire HERA data sample. 
The  covered kinematic range is $60<W<220$~GeV~and $Q^2<1$ GeV$^2$. The measurement of $b$, shown in fig.~\ref{fig4} (top), yielded
$b=4.3^{+2.0}_{-1.3}$ (stat.)$\, ^{+0.5}_{-0.6}$ (syst.)~GeV$^{-2}$, consistent with predictions based on pQCD models ($b=3.68$~GeV$^{-2}$)~\cite{DiffY}.
The result is in agreement with expectations of an asymptotic behavior of the slope parameter as 
a function of the scale and extends the scale range to $\sim 90$~GeV$^2$, the highest value achieved so far for a vector meson.

In fig.~\ref{fig4} (bottom) the already discussed  H1 preliminary results for $J/\psi$ are presented as differential cross-section $d\sigma/dt$ for elastic photo-production, together with the results of the exponential fit, for two energy values, $\sqrt{s}=318$~GeV and $\sqrt{s}=225$~GeV, corresponding to an higher and lower $W$ region, respectively.

\section{Exclusive two-pion production in DIS}

\begin{figure}[h]
\centering
 \includegraphics[scale=.33]{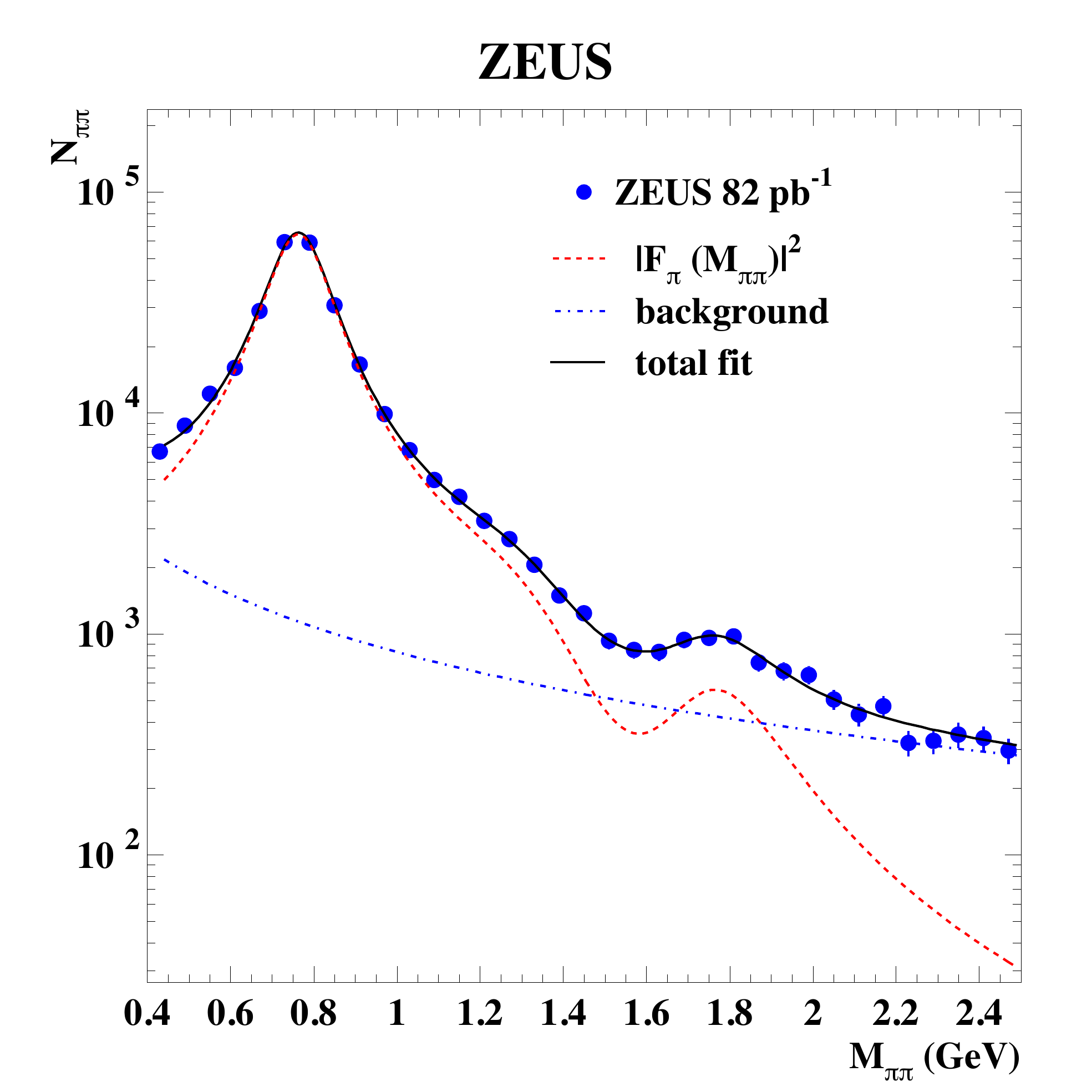}
\includegraphics[scale=.51]{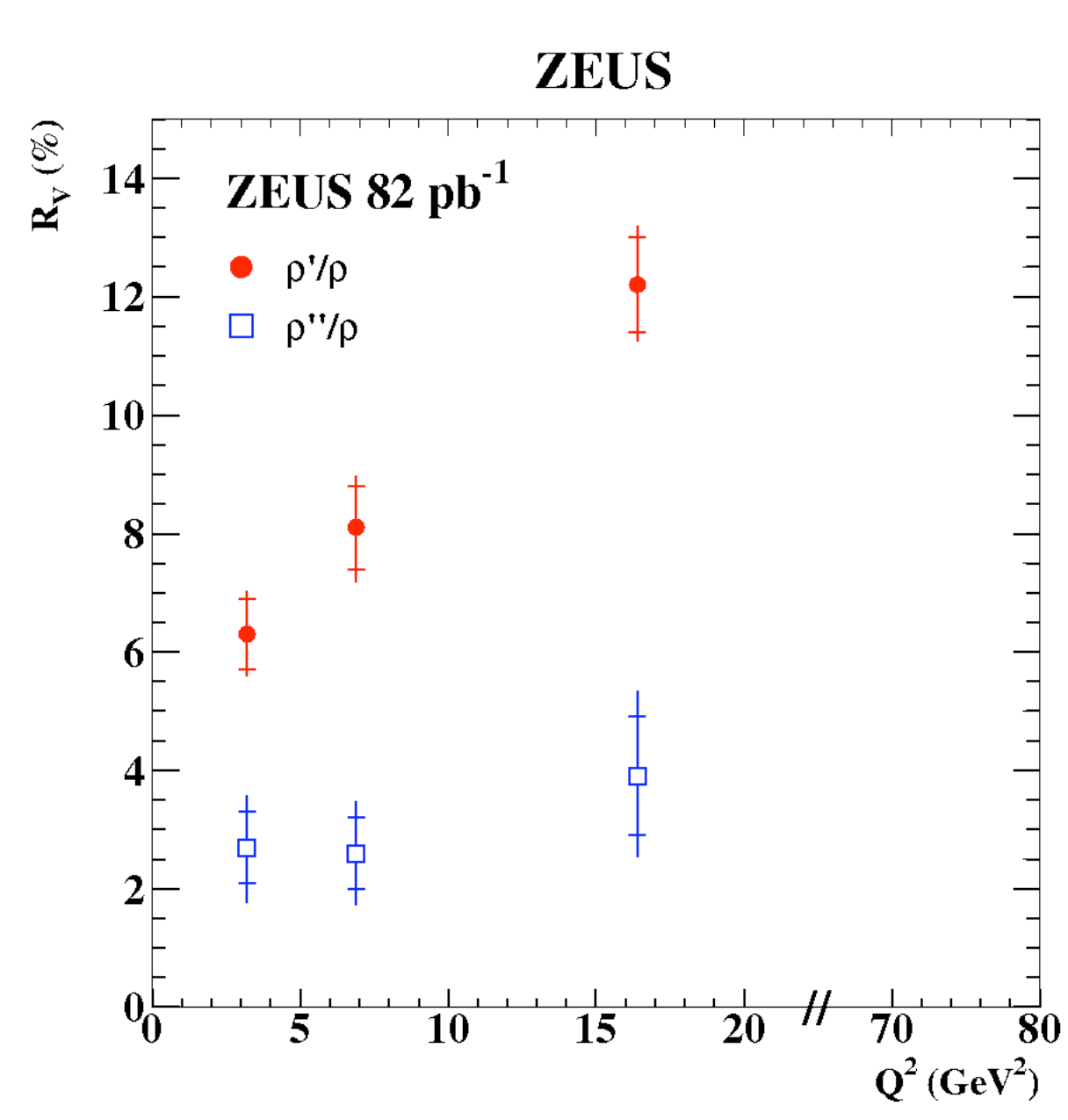}
\caption{(left) The two-pion invariant-mass distribution, $M_{\pi\pi}$, where 
$N_{\pi\pi}$ is the acceptance corrected number of events. The full line is the result of a
fit using the Kuhn-Santamaria parameterization~\cite{kuhn}. The dashed line is the
result of the pion form factor normalized to the data and the
dash-dotted line denotes the background contribution. (right) The ratio $R_V$ as a function of $Q^2$ for $V=\rho'$ (full circles) and $V=\rho''$ (open squares).}
\label{fig5}
\end{figure}

A new interesting result, presented by ZEUS, is the exclusive DIS production of two pions~\cite{twopi} in the mass range $0.4<M_{\pi\pi}<2.5$~GeV. The  measurement is performed in the kinematic range $2<Q^2<~80$~GeV$^2$, $32<W<180$~GeV  and $t<0.6$~GeV$^2$ and is shown in fig.~\ref{fig5} (left) where the two-pion acceptance-corrected data are compared with the pion electromagnetic form factor, $|F(M_{\pi\pi})|$, assuming that the mass range includes the contribution of the $\rho$, $\rho\rq{}(1450)$ and $\rho\rq{}\rq{}(1700)$ vector meson states.

The $Q^2$ dependence of the pion form factor has been also investigated. 
The data sample has been divided in three $Q^2$ bins and a fit to $M_{\pi\pi}$ performed. 
Figure~\ref{fig5} (right) shows the ratios $R_{\rho'}=\sigma(\rho\rq{} \rightarrow \pi\pi)/ \sigma(\rho)$ and $R_{\rho''}=\sigma(\rho\rq{} \rq{}\rightarrow \pi\pi)/ \sigma(\rho)$ as a function of $Q^2$. 
We observe a strong $Q^2$ dependence of the ratio  $R_{\rho'}$, as expected in
QCD-inspired models~\cite{twopiratio}. No conclusions are possible regarding the $Q^2$ dependence of  $R_{\rho'}$ due to the large uncertainties. 

\begin{figure}[h]
\centering
\includegraphics[scale=.6]{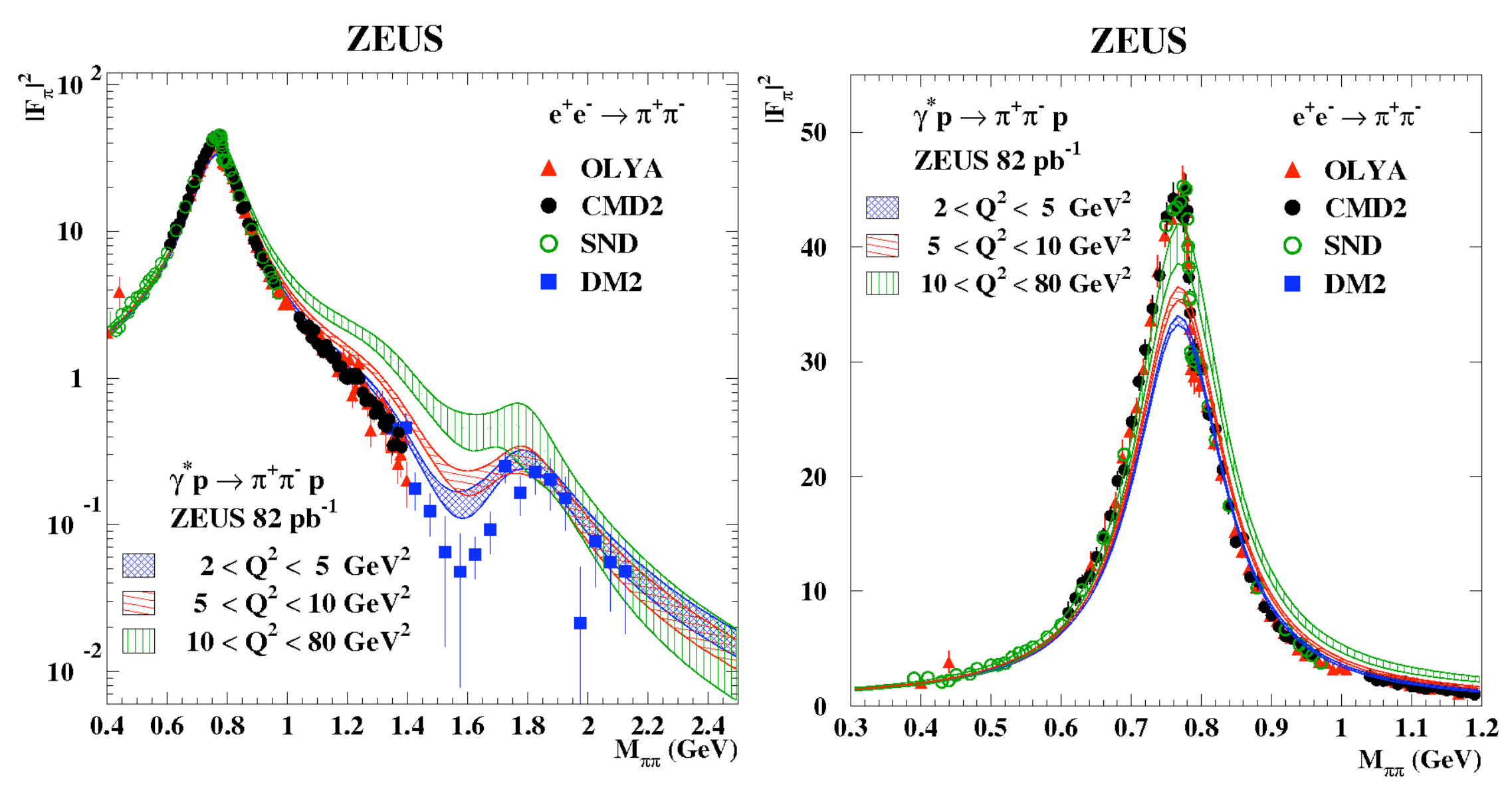}
\caption{(left) The pion form factor squared, $|F_{\pi}|^2$, as a function of 
the $\pi^+\pi^-$ invariant mass, $M_{\pi\pi}$, as obtained from the
reaction $e^+e^-\to\pi^+\pi^-$. The
shaded bands represent the square of the pion form factor and its total
uncertainty obtained in the $\gamma^*p$  analysis for three ranges of
$Q^2$. (right) The $\rho$ mass region of the left-hand-side figure is shown in a linear scale.}
\label{fig6}
\end{figure}

The last result presented, shown in 
Figure~\ref{fig6}, is the  pion form factor, $|F_{\pi}|^2$, resulting from the fit for the three $Q^2$ regions and measurements, in the time-like regime, of $e^+e^-\rightarrow \pi^+ \pi^-$~\cite{barkov,dm2,cmd2,cmd2-b,snd}.
A $Q^2$ dependence of $|F_\pi(M_{\pi\pi})|^2$ is observed. In particular, it is visible that in the interference region between $\rho\rq{}$ and $\rho\rq{}\rq{}$ (left) the distribution of the $\gamma^*p$ data is higher than the 
$e^+e^-$ data and lower in the $\rho$ mass range (right). The measurements are compatible for $M_{\pi\pi} > 1.8$ GeV.

\newpage

\begin{footnotesize}

\end{footnotesize}

\end{document}